\documentclass[fleqn,usenatbib]{mnras}

\usepackage{newtxtext,newtxmath}
\usepackage{mathptmx}
\usepackage{txfonts}

\usepackage[T1]{fontenc}
\usepackage{ae,aecompl}

\usepackage{graphicx}	
\usepackage{amsmath}	
\usepackage{amssymb}	
\usepackage{xcolor}
\usepackage{kotex}
\usepackage{siunitx}
\usepackage{soul,xspace}
\usepackage{array,multirow}
\usepackage{ulem}
\hypersetup{draft}

\newcommand {\nhunit} {cm$^{-2}$\xspace}
\newcommand {\nh} {$N_{\mathrm{H}}$\xspace}
\def\Plus{\texttt{+}}
\def\Minus{\texttt{-}}

\makeatletter
\newcommand*{\rom}[1]{\expandafter\@slowromancap\romannumeral #1@}
\makeatother

\title[Parsec-scale Jet Properties of BASS AGN]{BAT AGN Spectroscopic Survey - \rom{17}: The Parsec-scale Jet Properties of the Ultra Hard X-ray Selected Local AGN}

\author[J. Baek et al.]{
Junhyun Baek$^{1}$\thanks{E-mail: jhbaek1001@yonsei.ac.kr},
Aeree Chung$^{1}$\thanks{E-mail: achung@yonsei.ac.kr},
Kevin Schawinski$^{2}$,
Kyuseok Oh$^{3,4}$,
O. Ivy Wong$^{5}$,
\newauthor
\space Michael Koss$^{6}$,
Claudio Ricci$^{7,8}$,
Benny Trakhtenbrot$^{9}$,
Krista Lynne Smith$^{10,11}$,
\newauthor
\space and Yoshihiro Ueda$^{3}$
\\
\\
$^{1}$Department of Astronomy, Yonsei University, 50 Yonsei-ro, Seodaemun-gu, Seoul 03722, Korea\\
$^{2}$Institute for Particle Physics and Astrophysics, ETH Z\"urich, Wolfgang-Pauli-str. 27, CH-8093 Z\"urich, Switzerland\\
$^{3}$Department of Astronomy, Kyoto University, Oiwake-cho, Sakyo-ku, Kyoto 606-8502, Japan\\
$^{4}$JSPS fellow\\
$^{5}$International Centre for Radio Astronomy Research-M468, The University of Western Australia, 35 Stirling Hwy., Crawley, WA 6009, Australia\\
$^{6}$Eureka Scientific Inc., 2452 Delmer St. Suite 100, Oakland, CA 94602-3017, USA\\
$^{7}$Instituto de Astrofsica, Facultad de Fsica, Pontificia Universidad Cat\`{o}lica de Chile, Casilla 306, Santiago, Chile\\
$^{8}$Kavli Institute for Astronomy and Astrophysics, Peking University, Beijing 100871, China\\
$^{9}$School of Physics and Astronomy, Tel Aviv University, Tel Aviv 69978, Israel\\
$^{10}$Kavli Institute for Particle Astrophysics and Cosmology, Stanford University, SLAC National Accelerator Laboratory, Menlo Park, CA 94025, USA\\
$^{11}$Einstein Fellow
}

\date{Accepted 2019 July 15. Received 2019 July 14; in original form 2019 May 13}

\pubyear{2019}

\begin{document}
\label{firstpage}
\pagerange{\pageref{firstpage}--\pageref{lastpage}}
\maketitle

\begin{abstract}
We have performed a very long baseline interferometry (VLBI) survey of local ($z<0.05$) ultra hard X-ray (14$-$195 keV) selected active galactic nuclei (AGN) from the {\it Swift} Burst Alert Telescope (BAT) using KVN, KaVA, and VLBA. We first executed fringe surveys of 142 BAT-detected AGN at 15 or 22 GHz. Based on the fringe surveys and archival data, we find 10/279 nearby AGN ($\sim$4\%) VLBI have 22 GHz flux above 30\,mJy. This implies that the X-ray AGN with a bright nuclear jet are not common. Among these 10 radio-bright AGN, we obtained 22 GHz VLBI imaging data of our own for four targets and reprocessed archival data for six targets. We find that, although our 10 AGN observed with VLBI span a wide range of pc-scale morphological types, they lie on a tight linear relation between accretion luminosity and nuclear jet luminosity. Our result suggests that a powerful nuclear radio jet correlates with the accretion disc luminosity. We also probed the fundamental plane of black hole activity at VLBI scales (e.g., few milli-arcsecond). The jet luminosity and size distribution among our sample roughly fit into the proposed AGN evolutionary scenario, finding powerful jets after the blow-out phase based on the Eddington ratio ($\lambda_{\rm Edd}$)-hydrogen column density ($N_{\rm H}$) relation. In addition, we find some hints of gas inflow or galaxy-galaxy merger in the majority of our sample. This implies that gas supply via tidal interactions in galactic scale may help the central AGN to launch a powerful parsec-scale jet.

\end{abstract}




\section{INTRODUCTION}

\noindent Supermassive black holes (SMBHs) can not only accrete matter but also power strong jets along which material can be ejected \citep{Urry1995}. The synchrotron jets associated with active galactic nuclei (AGN) can deliver enormous kinetic energy to the surroundings, causing a wide range of feedback processes including both positive/negative effects on star formation activities \citep[e.g.,][]{Fabian2012, Harrison2017}. Therefore the AGN jet is believed to be an important component driving galaxy evolution \citep[e.g.,][]{Gaibler2012, Wagner2012, Zinn2013}.

The relation between synchrotron jet properties and the SMBH mass or accretion rate may be the key to understanding the AGN feedback process. In a simplified black hole model, the radiation is expected to be directly proportional to the accreting mass ($L$\,=\,$\varepsilon \overset{.}{M}c^{2}$; where $\varepsilon$ is the radiative efficiency), which is also commonly adopted in many cosmological hydrodynamic simulations where AGN feedback is implemented (e.g., EAGLE: \citealt{Schaye2015}; Horizon-AGN: \citealt{Dubois2012}; IllustrisTNG: \citealt{Springel2005}). This relation has also been confirmed by a number of observational studies, which probed the optical or soft X-ray brightness together with radio properties, mostly in low frequencies such as 1.4 GHz or 5 GHz \citep[e.g.,][]{Allen2006, Panessa2007, Panessa2015, Sikora2007}. One of the most interesting relationships is the correlation between accretion luminosity, black hole (BH) mass and jet luminosity, the so-called `fundamental plane of black hole activity’, which seems to hold from stellar mass BHs to SMBHs \citep{Merloni2003, Falcke2004}.

Very long baseline interferometry (VLBI) observations at high radio frequencies of a few tens GHz is a powerful tool to measure the brightness of nuclear synchrotron jets associated with SMBHs. Thanks to the extremely high resolution (few milli- and sub milli-arcseconds scale) provided by VLBI observations, we can also study the morphology of nuclear radio emission with few sub-milli arcsec resolutions, which corresponds to sub-parsec scales at $z<0.05$. At high radio frequencies ($\nu$ of a few GHz), the synchrotron self-absorption or the free-free absorption is not as significant as in low frequencies \citep{Kellermann1966}. Radio activities from the nuclear region of recently powered AGN can then be probed even when embedded in the dense ambient medium of the nuclear region \citep[e.g.,][]{Snellen1999,Tinti2005,Jeong2016}. 

Rapid accretion onto the SMBH is expected to accompany the formation of powerful jets, so the accretion mass estimated from optical or X-ray brightness should show a correlation with the jet luminosity. This relation, however, seems to be valid only for extended kpc-scale jets, but not for pc-scale nuclear jets \citep[e.g.,][]{Panessa2013}. The lack of correlation is surprising as nuclear jets trace recent accretion events while kpc-scale jets are representative of past accretion. However, the previous studies of the pc-scale jet and other BH properties have not covered a broad range of AGN populations. For example, \citet{Panessa2013} studied 28 Seyfert galaxies in the very nearby Universe ($D_{\rm L}$ < 23 Mpc, $z<0.0054$), which limited the sample to low luminosity AGN (LLAGN). In LLAGN/radio-quiet AGN the origin of radio emission could also be from bremsstrahlung free-free accretion disc winds or non-thermal corona, i.e. a hot Comptonized envelop in the vicinity of the accretion disc \citep{Haardt1991, Christopoulou1997, Panessa2019}. A broader sample including more luminous AGN is critical to understand the relationship between nuclear jets and SMBHs. 

AGN found in the hard X-rays, where photons are energetic enough to pass through the dense gas and dust in the vicinity of BHs, are less biased to obscuration compared to optical/UV selected samples \citep{Mushotzky1976}.  The hard X-rays are also emitted primarily by the inverse Compton effect of thermal optical/UV photons from the accretion disc \citep{Haardt1993}, so the luminosity of the accretion disc can be directly compared with nuclear jet properties.

In this work, we present the result of our 22 GHz VLBI study of nearby ($D_{\rm L}$ < 220 Mpc or $z<0.05$) X-ray selected AGN sample from the {\it Swift}-Burst Alert Telescope (BAT) survey including LLAGN to more luminous AGN. To date, several studies have investigated the radio properties of the BAT AGN in various frequencies (e.g. 1.4 GHz: \citealt{Wong2016}; 22 GHz: \citealt{Burlon2013, Smith2016}; Smith et al. submitted). Those studies focused on the kpc-scale radio properties of the sample, whereas this work is the first systematic study probing the pc-scale nuclear structure of the BAT AGN using the VLBI facilities.

This paper is organized as follows. In Section 2, we introduce the sample and our initial study of archival 5 GHz data to identify potential targets for VLBI imaging. In Section 3, we describe the fringe surveys and the VLBI imaging observations and data reduction. In Section 4, we present the results, which are further discussed and summarized in Sections 5 and 6, respectively. In this work, we adopt a standard cosmology of $\Omega_{\Lambda}$ = 0.7, $\Omega_{M}$ = 0.3, and $H_{0}$ = 70 km s$^{-1}$ Mpc$^{-1}$.

\section{SAMPLE}
\label{sec:sample}

\begin{table}
	\centering
	\caption{Overview of sample selection in this study}
	\label{tab:tab1}
	\begin{tabular}{cc} 
		\hline
		\hline
		Selection criterion & The number of AGN\\
		\hline
		\hline
		{\it Swift}-BAT 70-month catalog & 836\\
		BASS DR1 & 641\\
		$z<0.05$ & 377\\
		Decl. $>\ang{-30}$ & 283\\
		Non-blazars & 279\\
		\hline
		GB6/PMN cross-matched source & 47\\
		KaVA observations \& VLBA archive data & 10\\
		\hline
		\hline
	\end{tabular}
\end{table}

\noindent Our parent sample comes from the {\it Swift}-BAT all sky survey conducted in the 14-195 keV energy band with 1.34$\times$10$^{-11}$ ergs s$^{-1}$ cm$^{-2}$ sensitivity limit \citep{Baumgartner2013}. The {\it Swift}-BAT 70-month catalog contains 836 AGN identified in the survey (c.f. 1099 AGN identified by the 105-month catalog, \citealt{Oh2018}), providing an AGN sample that is unbiased to obscuration up to Compton-thick levels (\nh$<10^{24}$ \nhunit; \citealt{Burlon2011, Ricci2015}).

\begin{table*}
	\centering
	\caption{Summary of the fringe surveys}
	\label{tab:tab2}
	\resizebox{0.75\textwidth}{!}{
	\begin{tabular}{cccccc} 
		\hline
		\hline
		Frequency & Telescope & The number of & Detections & Non-detections & Detection limit\\
		 & & observed targets & & & (5$\sigma$ level) \\
		\hline
		\hline
		\multicolumn{6}{l}{\emph{47 GB6/PMN cross-matched sources at 5 GHz}} \\
		22 GHz & KVN & 31 & 2 & 29 & 30.5\,mJy$^{*}$ \\
		22 GHz & VLBA archive & 16 & 8 & 8 & 16.3\,mJy$^{*}$ \\
		\hline
		\multicolumn{6}{l}{\emph{Blind fringe survey in RA 08h$-$12h and 21.5h$-$02.5h}}\\
		15 GHz & VLBA & 95 & 0 & 95 & 13.4\,mJy$^{*}$ \\
		\hline
		\multicolumn{6}{l}{\emph{In total}} \\
		 & & 142 & 10 & 132 \\
		\hline
		\hline
	\end{tabular}
	}
	\begin{flushleft}
	\hspace{2.4cm} $^{*}$ The baseline sensitivity (1$\sigma$) in 120 seconds of integration over 256 MHz bandwidth.
	\end{flushleft}
\end{table*}

The sample for our VLBI study has been selected from 836 AGN following with these criteria: (1) the target should have obscuration corrected X-ray luminosity from \citet{Ricci2017b} and a black hole mass from the BAT AGN Spectroscopic Survey (BASS)\footnote{https://www.bass-survey.com} Data Release 1 \citep[DR1;][]{Koss2017}, (2) it needs to be at $z<0.05$ so that parsec-scale structures are fairly well resolved, (3) its declination should be >~$\ang{-30}$ to be accessible with the Very Long Baseline Array (VLBA) and the Korean VLBI Network (KVN), and (4) it should not be classified as a blazar in order to avoid a Doppler boosting effect and hence an overestimation of jet brightness. BL Lac objects (BZB class) and the flat spectrum radio quasars (BZQ class) identified in the 5th edition of the ``Multifrequency Catalogue of BLAZARS (Roma-BZCAT)\footnote{https://www.asdc.asi.it/bzcat/}" \citep{Massaro2015} are left out from our sample. Table~\ref{tab:tab1} lists the number of targets remained in the sample after each criterion is applied from (1) to (4). This left 279 targets for our radio VLBI study.

In order to select AGN bright enough for VLBI imaging, we used the single-dish 5 GHz flux. This is the highest radio frequency at which almost the entire sky has been surveyed with uniform sensitivity (18$-$42\,mJy) using the 100 m Green Bank Telescope (Green Bank 6-cm Survey: GB6; \citealt{Gregory1996}) and the 64 m Parkes telescope (Parkes-MIT-NRAO survey: PMN; \citealt{Griffith1993}) in the northern and the southern hemisphere, respectively. For a typical AGN with a synchrotron power-law spectrum ($S_{\nu}$ $\varpropto$ $\nu^{-\alpha}$; $\alpha \sim$ 0.7), only the ones detected in the GB6/PMN survey are likely to be visible at 22 GHz using the KaVA (the combined array of KVN and VERA) for which the 5$\sigma$ baseline sensitivity is 30\,mJy.

When cross-matched with the GB6/PMN catalog within a 105-arcsec radius (the primary beam size of the survey), 47/279 (17\%) AGN are detected, providing potential targets for our VLBI imaging. Of these 47 AGN (listed in Table~\ref{tab:tabA1}), 16 AGN already had archival VLBA imaging at 22 GHz, with 8 detections.

\section{OBSERVATIONS AND DATA REDUCTION}

\noindent We first performed a KVN fringe survey at 22 GHz to verify the VLBI detectability for the 31/47 AGN identified from the 5 GHz flux as possible bright enough for VLBI imaging, but without existing images. Additional 2 out of 31 AGN were detected in our KVN fringe survey, hence the total number of AGN to be imaged at 22 GHz was 10 in total. We then performed a VLBA blind fringe survey at 15 GHz of 95 BAT AGN in two specific RA ranges (08h$-$12h and 21.5h$-$02.5h) and identified no additional sources beyond the 10 AGN already identified using the fringe survey or archival data. The number of targets in the fringe surveys is summarized in Table~\ref{tab:tab2}. Of these 10 AGN bright at 22 GHz, six already had archival high-quality VLBA imaging which we reprocessed, two (NGC 1275 and PKS 2331-240) we obtained better quality KaVA data, and we also obtained KaVA imaging data for two sources for the first time (see Table~\ref{tab:tab3}).

\begin{figure}
	\includegraphics[width=\columnwidth]{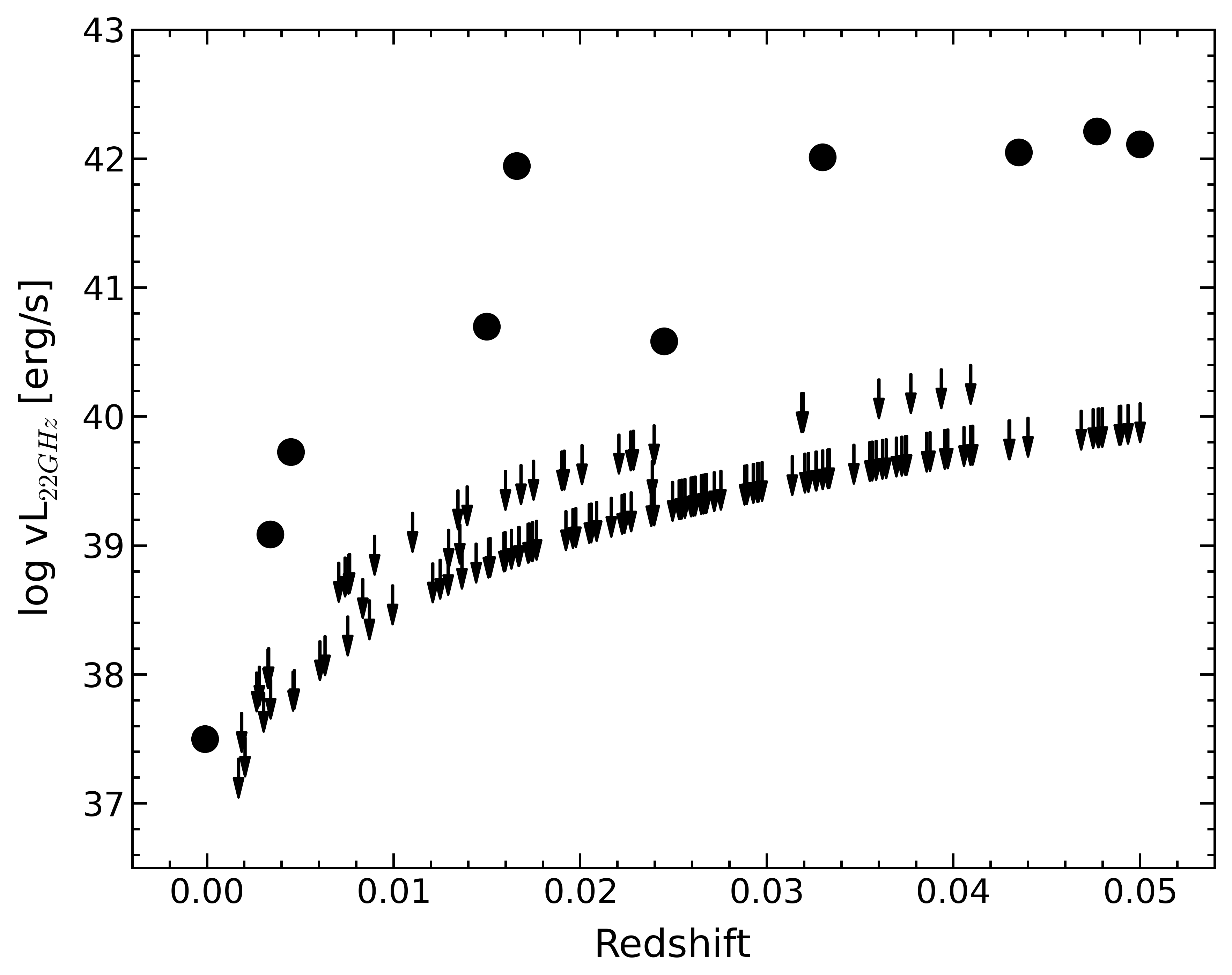}
    \caption{The 22 GHz luminosity of 142 targets observed using the KVN and the VLBA is shown as a function of redshift. Big dots indicate 10 detections and downward arrows are the upper limits of non-detections. The upper limits are 30.5 mJy for the KVN observation and 16.3 mJy for the VLBA archival data. The upper limit for our own VLBA at 22 GHz was estimated from the upper limit of our 15 GHz observation, assuming $S_{\nu}$ $\varpropto$ $\nu^{-0.7}$. The sample of 10 VLBI-detected BASS AGN has a broad range of radio luminosity including LLAGN, but the majority are luminous AGN.}
    \label{fig:figure1}
\end{figure}

\begin{table*}
	\centering
	\caption{22GHz VLBI imaging observations}
	\label{tab:tab3}
	\begin{tabular}{ccccccccc} 
		\hline
		\hline
		BAT & Name & Telescope & Obs.code & Obs.date & Bandwidth & Integ.time & Resolution & $S_{\rm 22GHz,VLBI}$ \\
		index & & & & (yyyy.mm.dd) & (MHz) & (minutes) & (mas $\times$ mas, deg) & (Jy) \\
		(1) & (2) & (3) & (4) & (5) & (6) & (7) & (8) & (9) \\
		\hline
		\hline
		173 & NGC 1275 & KaVA & K15JB01A & 2015.10.13 & 256 & 25 & 1.07 $\times$ 0.99, 80.8 & 8.618 $\pm$ 1.293 \\
		579 & NGC 3998 & KaVA & K17JB01A & 2017.03.25 & 256 & 105 & 1.37 $\times$ 0.99, 78.9 & 0.127 $\pm$ 0.019 \\
		876 & ARP 102B & KaVA & K17JB01A & 2017.03.25 & 256 & 150 & 1.14 $\times$ 0.97, 37.3 & 0.173 $\pm$ 0.026 \\
		1200 & PKS 2331-240 & KaVA & K17JB01A & 2017.03.25 & 256 & 90 & 1.88 $\times$ 0.99, 7.6 & 1.902 $\pm$ 0.285 \\
		\hline
		33 & Mrk 348 & VLBA & BP182 & 2014.04.20 & 256 & 80 & 0.96 $\times$ 0.28, -6.8 & 0.461 $\pm$ 0.069 \\
		140 & NGC 1052 & VLBA & BR130H & 2009.03.08 & 64 & 57 & 0.84 $\times$ 0.33, 5.6 & 0.541 $\pm$ 0.081 \\
		147 & Q0241+622 & VLBA & BH136F & 2006.11.29 & 32 & 36 & 0.69 $\times$ 0.25, 9.4 & 1.212 $\pm$ 0.182 \\
		214 & 3C 111.0 & VLBA & BT104 & 2009.12.18 & 32 & 33 & 0.52 $\times$ 0.32, -8.6 & 1.057 $\pm$ 0.158 \\
		226 & 3C 120 & VLBA & BG182B & 2007.11.07 & 32 & 165 & 1.27 $\times$ 0.39, -19.6 & 1.939 $\pm$ 0.291 \\
		477 & M 81 & VLBA & BH173 & 2001.10.03 & 128 & 32 & 0.36 $\times$ 0.32, 24.7 & 0.088 $\pm$ 0.013 \\
		\hline
		\hline
	\end{tabular}
	\begin{flushleft}
	Note. (1) Entry number listed in the {\it Swift} BAT 70-month hard X-ray survey \citep{Baumgartner2013}; (2) Source name; (3) VLBI name; (4) Observation code; (5) Observation date; (6) Observation bandwidth in MHz; (7) On-source time in minutes; (8) Synthesized beam size of image; (9) Observed total flux density at 22 GHz.
	\end{flushleft}
\end{table*}

\subsection{KVN and VLBA fringe survey}

In order to select the targets that are detectable with the VLBI, we first carried out a KVN fringe survey at 21.7 GHz on 24$-$25 March 2016. In total, 31 out of 47 targets were observed for 5 minutes $\times$ 3 times, at different azimuths in order to avoid intermittently occurring bad weather and increase the chance for detection. For 16 sources with publicly available VLBA data, we used the archival data to estimate the brightness. In the KVN fringe survey, a Mark 5B recorder was used at a recording rate of 1 Gbps in single polarization of a bandwidth of 256 MHz with 2-bit quantizations and Nyquist sampling. The 256 MHz bandwidth was divided into 8 IFs of 32 MHz with 128 channels for each IF. The data were correlated with the DiFX software correlator in the Korea-Japan Correlator Center \citep{Lee2015a}.

A standard post-correlation process was performed with the NRAO Astronomical Image Processing System ({\tt AIPS}; \citealt{Fomalont1981}; {\tt 31DEC15 version}). A possible digital sampling loss on the amplitudes of cross-correlation spectra were corrected by the task {\tt VLBACCOR}. The instrumental phase of each baseline was manually corrected by the task {\tt VLBAMPCL}. The task {\tt VLBABPSS} was used for determining antenna bandpass functions and scaling the amplitude with auto-correlations. Using the antennas' gain curves and opacity-corrected system temperatures provided by the observatory, a-priori amplitude corrections were applied by the task {\tt VLBAAMP}. Any uncorrected residuals of delay and rate were corrected by fringe fitting using {\tt FRING} and {\tt CLCAL}. Based on our KVN fringe survey and the VLBA archival data, we found 10 detectable sources at 22 GHz for the follow-up VLBI imaging.

The sample for our initial fringe survey using the KVN was selected based on the assumption of a power-law spectrum of AGN. In some cases, a young AGN with a slope deviating from our assumption \citep[e.g., a convex spectrum;][]{Tinti2005} might have been missed from our selection can be still detected at 22 GHz, but may have no counterpart at 5 GHz.  To overcome this issue and estimate sampling bias, we conducted a 15 GHz blind VLBA fringe survey on 279 sources which passed our initial four selection criteria (BB395; PI: J. Baek). With limited observing time, we covered 95 objects in two RA ranges, 08h$-$12h and 21.5h$-$02.5h ($\sim$37.5\% of all sky above Decl. -30 deg), none of which overlaps with the sample of 47 AGN that we already had 22 GHz VLBI data for.

The VLBA fringe survey at 15.4 GHz was carried out on 16, 19, 24 March 2018 for total 11 hrs. The on-source time of each target is 5 minutes with 10 antennas, which guarantee the high chance for detection within single scan regardless of local or intermittent bad weather. A Mark 5C recorder was used at recording rate of 2 Gbps. The 256 MHz bandwidth in each polarization was divided into 2 IFs of 128 MHz with 256 channels of each IF. The data were correlated with the Socorro-DiFX correlator \citep{Deller2011}. The correlated VLBA data were reduced with the same {\tt AIPS} procedures with previous KVN fringe survey.

With these 95 observations (listed in Table~\ref{tab:tabA2}), we did not find any new potential target for the VLBI imaging. Although our combined fringe surveys and archival data only covered 142/279 AGN ($\sim$51\%), the lack of detection of any of the 95 AGN in the blind survey combined with the 5 GHz archival data suggests that only 4\% (10/279) of the local, ultra hard X-ray selected AGN are bright enough to be imaged in high radio frequency at the pc-scale.

Figure~\ref{fig:figure1} shows the luminosity distribution of the 142 VLBI observed targets. The 10 VLBI-detected BASS AGN sample spans 10$^{37}$ $-$ 10$^{42.5}$ erg s$^{-1}$ in the nuclear radio luminosity, where we do not expect faint radio emission originating from the corona, disc winds, and/or central star formation. Hence our sample consists of radio bright AGN population compared with the previous low-frequency VLBI studies \citep[e.g.,][]{Panessa2013} or same frequency VLA deep imaging study (\citealt{Smith2016}; Smith et al. submitted).

\begin{figure*}
	\includegraphics[width=\textwidth]{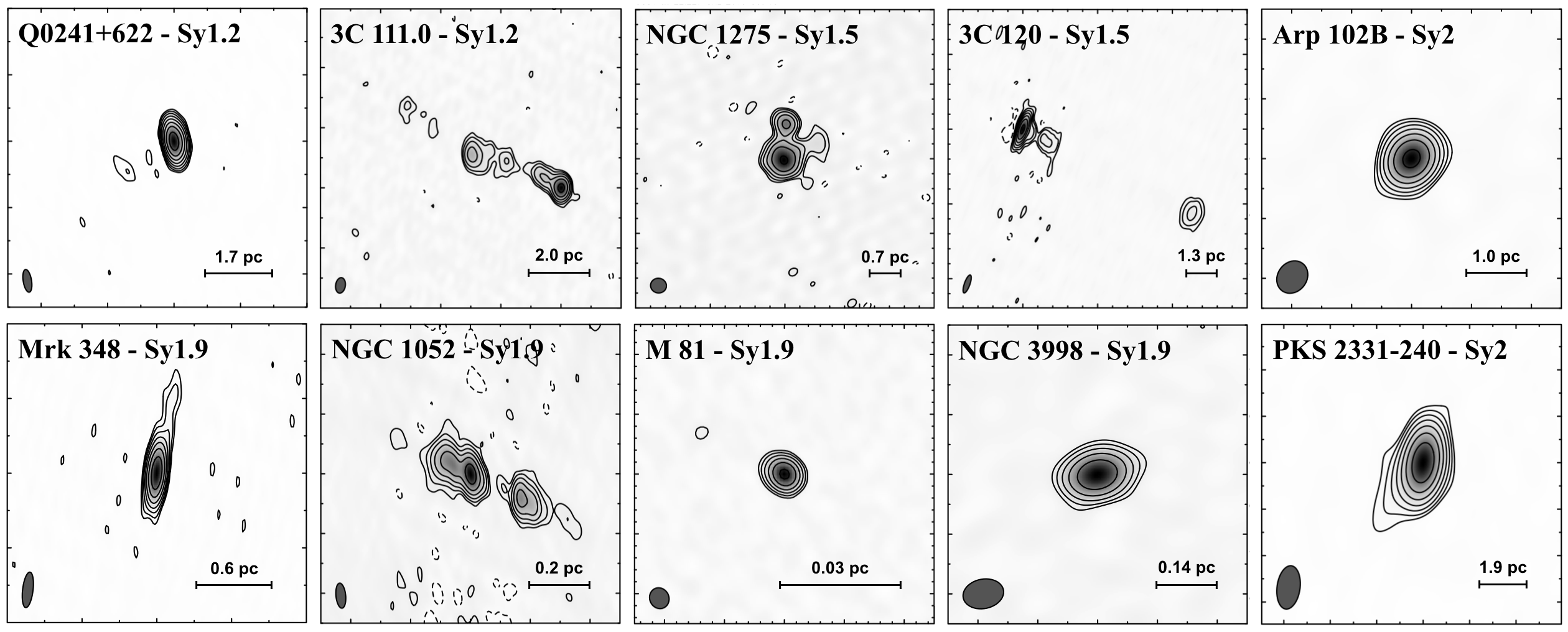}
    \caption{VLBI 22 GHz parsecc-scale radio morphologies of our sample. On the right-hand side of the target name, the Seyfert type is presented. Depending on the image quality, the lowest contour was drawn at 3 to 5 $\times$ r.m.s. (3$\sigma$ - 3C 111.0, NGC 1275, 3C 120, Arp 102B, NGC 1052; 4$\sigma$ - Q0241+622, M 81, NGC\,3998, PKS 2331-240; 5$\sigma$ - Mrk 348), and the rest increases by $\times$ 2$^{\rm n}$. The synthesized beam is shown in the bottom left corner of each panel. The bar in the bottom right corner of each panel is 2 milli-arcsec in size, which corresponds to 0.03 $\sim$ 2 pc depending on the distance to the target. A broad range of radio morphology is found among our VLBI sample in (sub)pc-scale.}
    \label{fig:figure2}
\end{figure*}

\begin{table*}
	\centering
	\caption{22 GHz VLBI luminosity, intrinsic ultra hard X-ray luminosity, and black hole mass of our VLBI sample}
	\label{tab:tab4}
	\begin{tabular}{ccccccccc} 
		\hline
		\hline
		BAT & Name & log $\nu L_{\rm 22 GHz, VLBI}$ & log $L_{\rm 14-195 keV}$ & log $L_{\rm bol}$ & log $M_{\rm BH}$ & Method & ${\rm log}~\lambda_{\rm Edd}$ & ${\rm log}~N_{\rm H}$ \\
		index & & (erg s$^{-1}$) & (erg s$^{-1}$) & (erg s$^{-1}$) & ($M_{\odot}$) & & & (cm$^{-2}$) \\
		(1) & (2) & (3) & (4) & (5) & (6) & (7) & (8) & (9) \\
		\hline
		\hline
		\vspace{1mm}
		33 & Mrk 348 & 40.698 $\pm$ 0.065 & 43.90$^{\Plus0.05}_{\Minus0.08}$ & 44.77$^{\Plus0.05}_{\Minus0.08}$ & 7.61 $\pm$ 0.64 & $M_{\rm BH}$-$\sigma_{*}^{1}$ & -0.95$^{\Plus0.69}_{\Minus0.72}$ & 23.12$^{\Plus0.03}_{\Minus0.02}$ \\
		\vspace{1mm}
		140 & NGC 1052 & 39.725 $\pm$ 0.065 & 42.18$^{\Plus0.02}_{\Minus0.14}$ & 43.05$^{\Plus0.02}_{\Minus0.14}$ & 8.96 $\pm$ 0.29 & $M_{\rm BH}$-$\sigma_{*}^{(1)}$ & -4.02$^{\Plus0.31}_{\Minus0.43}$ & 22.95$^{\Plus0.02}_{\Minus0.01}$ \\
		\vspace{1mm}
		147 & Q0241+622 & 42.048 $\pm$ 0.065 & 44.71$^{\Plus0.06}_{\Minus0.17}$ & 45.58$^{\Plus0.06}_{\Minus0.17}$ & 8.09 $\pm$ 0.10 & broad H$\beta^{(1)}$ & -0.62$^{\Plus0.16}_{\Minus0.27}$ & 20.92$^{\Plus0.08}_{\Minus0.09}$ \\
		\vspace{1mm}
		173 & NGC 1275 & 42.058 $\pm$ 0.065 & 43.75$^{\Plus0.07}_{\Minus0.03}$ & 44.62$^{\Plus0.07}_{\Minus0.03}$ & 8.53 $\pm$ 0.18 & gas kinematics$^{2}$ & -2.02$^{\Plus0.25}_{\Minus0.21}$ & 21.68$^{\Plus0.10}_{\Minus0.06}$ \\
		\vspace{1mm}
		214 & 3C 111.0 & 42.111 $\pm$ 0.065 & 44.81$^{\Plus0.01}_{\Minus0.73}$ & 45.68$^{\Plus0.01}_{\Minus0.73}$ & 8.27 $\pm$ 0.10 & broad H$\beta^{(1)}$ & -0.70$^{\Plus0.11}_{\Minus0.83}$ & 21.87$^{\Plus0.08}_{\Minus0.11}$ \\ 
		\vspace{1mm}
		226 & 3C 120 & 42.011 $\pm$ 0.065 & 44.37$^{\Plus0.02}_{\Minus0.22}$ & 45.24$^{\Plus0.02}_{\Minus0.22}$ & 7.75 $\pm$ 0.04 & reverberation$^{(1,3)}$ & -0.62$^{\Plus0.06}_{\Minus0.26}$ & 21.45$^{\Plus0.13}_{\Minus0.19}$ \\
		\vspace{1mm}
		477 & M 81 & 37.499 $\pm$ 0.065 & 40.43$^{\Plus0.24}_{\Minus0.11}$ & 41.30$^{\Plus0.24}_{\Minus0.11}$ & 7.85 $\pm$ 0.11 & gas kinematics$^{(1,4)}$ & -4.66$^{\Plus0.35}_{\Minus0.22}$ & 19.95$^{\Plus0.09}_{\Minus0.17}$ \\
		\vspace{1mm}
		579 & NGC 3998 & 39.201 $\pm$ 0.065 & 41.62$^{\Plus0.02}_{\Minus0.25}$ & 42.49$^{\Plus0.02}_{\Minus0.25}$ & 8.91 $\pm$ 0.12 & stellar kinematics$^{(1,5)}$ & -4.53$^{\Plus0.14}_{\Minus0.37}$ & 20.66$^{\Plus0.15}_{\Minus0.32}$ \\
		\vspace{1mm}
		876 & ARP 102B & 40.701 $\pm$ 0.065 & 43.36$^{\Plus0.05}_{\Minus0.09}$ & 44.23$^{\Plus0.05}_{\Minus0.09}$ & 8.05 $\pm$ 0.10 & reverberation$^{6}$ & -1.94$^{\Plus0.15}_{\Minus0.19}$ & 21.32$^{\Plus0.13}_{\Minus0.17}$ \\
		1200 & PKS 2331-240 & 42.327 $\pm$ 0.065 & 43.86$^{\Plus0.08}_{\Minus0.05}$ & 44.73$^{\Plus0.08}_{\Minus0.05}$ & 8.75 $\pm$ 0.11 & $M_{\rm BH}$-$\sigma_{*}^{7}$ & -2.13$^{\Plus0.19}_{\Minus0.16}$ & 20.30$^{\Plus0.32}_{\Minus0.52}$ \\
		\hline
		\hline
	\end{tabular}
	\begin{flushleft}
	Note. (1) Entry number listed in the {\it Swift} BAT 70-month hard X-ray survey \citep{Baumgartner2013}; (2) Source name; (3) Measured 22 GHz VLBI luminosity with 15\% flux uncertainty; (4) Obsorption corrected intrinsic 14-195 keV luminosity from \citet{Ricci2017b}; (5) Bolometric luminosity inferred from $L_{\rm 14-195keV}$ in a factor of 7.42 (equivalent to the correction factor used in BASS DR1 if assuming $\Gamma$ = 1.8; \citealt{Koss2017}); (6) and (7) Black hole mass and derived methods, respectively. $^{(1)}$\citet{Koss2017}, $^{(2)}$\citet{Wilman2005}, $^{(3)}$\citet{Bentz2015}, $^{(4)}$\citet{Devereux2003}, $^{(5)}$\citet{Walsh2012}, $^{(6)}$\citet{Shapovalova2013}, $^{(7)}$\citet{Hernandez-Garcia2017}; (8) The logarithm of Eddington ratio derived by log($L_{\rm bol}$/$L_{\rm Edd}$); (9) Nutral Hydrogen column density from X-ray spectrum fitting \citep{Ricci2017b}.
	\end{flushleft}
\end{table*}

\subsection{KaVA and archive VLBA imaging}

\noindent To image parsec-scale AGN jets, we carried out KaVA follow-up observations for 4 out of 10 targets, at 21.8 GHz on 25$-$26 March 2017. All the instruments were configured in the same way as the KVN fringe survey, except for the number of antennas, which is 7 in the case of KaVA instead of 3 as for the KVN. To improve imaging quality, we spent $\sim$2 hours of on-source time on each target by taking multiple scans at different elevations, covering a wide range of uv-coverage. For 6 sources which were observed using the VLBA with sensitivity similar to or better than we could achieve with our KaVA observation, we used the VLBA data from the archive instead. The details of VLBI imaging observations can be found in Table~\ref{tab:tab3}.

For calibration, we again followed a standard post-correlation process using the {\tt AIPS (31DEC15 version)}. For the KaVA data there is amplitude loss due to the combination of 2-bit quantization in the digital filtering system and the characteristics of Daejeon correlator \citep{Lee2015b}, and hence we multiplied by 1.3 to recover the flux. Imaging of calibrated data was done using the DIFference MAPping software ({\tt DIFMAP}; \citealt{Shepherd1994}). Visibilities in all baselines with large errors in both phase and amplitude were flagged. A uniform weighting was applied to obtain the smallest possible synthesized beam to better resolve the nuclear jet structure. A loop of {\tt CLEAN}-{\tt SELFCAL}-{\tt GSCALE} was run iteratively to deconvolve point visibilities, to match the structure model, and to adjust the antenna gain, respectively.

\section{RESULTS}

\subsection{Parsec-scale morphologies and brightness}

\noindent The main interest of this work is the relationship between the accretion activity and the nuclear jet property of AGN. However, the major power sources of the radio emission in AGN (including LLAGN) can have three possible origins: 1) the non-thermal synchrotron radiation associated with jets, 2) the same but with corona, or 3) the thermal free-free emission from the accretion disc winds \citep[e.g.,][]{Begelman1984, Haardt1991,Christopoulou1997, Panessa2019}. Therefore, in order to probe the connection between accretion and jets, we first need to verify that the radio emission in our VLBI data mainly originates from the synchrotron jet.

The parsec-scale morphology revealed by VLBI observations can enable us to identify the origin of the nuclear source as the corona/wind powered emission is unlikely to show a jet-like structure. The jetted morphology of most of our targets have been already reported by previous VLBI studies. Mrk 348, NGC 1052 and NGC 1275 revealed nuclear jets in the past VLBA observations in multi-frequencies from 1.4 GHz up to 43 GHz with synchrotron spectra \citep[e.g.,][]{Walker2000, Kameno2001, Kadler2002, Peck2003, Vermeulen2003}. Q0241+622, 3C 111.0, and 3C 120 are also fairly well known for their nuclear jets as the sample of the Monitoring of jets in Active galactic nuclei with VLBA experiments (MOJAVE\footnote{https://www.physics.purdue.edu/MOJAVE/}), a long-term program which has been probing the jet kinematics at 15 GHz \citep[e.g.,][]{Chatterjee2009, Chatterjee2011, Lister2018}. In addition, PKS 2331-240 was found with a prominent nuclear jet structure in the previous VLBA observation at 5 GHz \citep{Fomalont2000}. We find that the results from our own observations and reprocessed archival data are consistent with those in the literature.

As seen in Figure~\ref{fig:figure2}, our sample shows a broad range of morphology at 22 GHz. Intriguingly, 7 out of 10 targets clearly reveal a (sub)pc-scale jet, which reconfirms that the synchrotron radiation is the main emitting source in radio wavelengths of these objects. Meanwhile, the remaining 3 targets are found with a compact core-like structure (M 81, NGC 3998, and Arp 102B). Among these, a pc-scale jet has been confirmed in M 81 in the previous VLBI study at 1.7 $-$ 8.4 GHz and 43 GHz \citep{Marti-Vidal2011, Ros2012}, which is also likely the origin of the radio emission.

For NGC 3998 and Arp 102B, we measured the brightness temperature ($T_{\rm b}$) of the radio emission to investigate the power source. In general, $T_{\rm b}$ of non-thermal emission is known to be higher ($T_{\rm b}$ > 10$^{7}$ K) than thermal emission ($T_{\rm b}$ $\sim$ 10$^{4}$ $-$ 10$^{5}$ K) \citep[e.g.,][]{Middelberg2004, Bontempi2012}. Using the extent of the nuclear radio component and its flux density, the brightness temperature can be estimated by the following equation:

\begin{center}
$T_{\rm b}$ (K) = 1.222 $\times$ 10$^{3}$ $\frac{I}{\nu^{2} \theta_{maj} \theta_{min}}$
\end{center}

\noindent where $I$ is the flux density in mJy, $\nu$ is the observing frequency in GHz, and $\theta$ is the size of the radio component in arcsec. For the unresolved compact sources, we measured the lower limit of brightness temperature using the synthesized beam size, yielding 10$^{8.4}$ K and 10$^{8.6}$ K for NGC 3998 and Arp 102B, respectively, i.e. sufficiently high to expect a non-thermal origin. Therefore we conclude that the radio emission detected in our VLBI image for these two cases also originates from non-thermal synchrotron radiation.

In AGN, the central synchrotron radiation can be still contaminated by the corona component. In the case where the corona emission is dominant (e.g. some radio-quite AGN or coronally active stars), the relative strength of radio to X-ray luminosity, $L_{\rm R}$/$L_{\rm X}$ is known to be $\sim$10$^{-5}$ \citep{Laor2008}. However, $L_{\rm R}$/$L_{\rm X}$ of our sample ranges from 10$^{-3.5}$ to 10$^{-1.5}$, which is comparable to jet-dominated, radio-loud AGN \citep{Laor2008}. Therefore our sample is likely to be jet-powered AGN where the synchrotron radiation is dominant.

\subsection{Fraction of AGN ejecting a nuclear jet}

\noindent A fundamental question of AGN feedback is what fraction of accreted mass into SMBH is ejected as a nuclear jet. Among the complete sample of 279 AGN in the local Universe ($z<0.05$), we find only $\sim$4\% (10 targets) with a bright extended jet in (sub)pc-scale. As shown in Figure~\ref{fig:figure1}, our sensitivity limit is sufficiently low compared to the brightness of the detections at high-z ($z\gtrsim0.03$). This implies that our low VLBI detection rate is likely to be real, and powerful nuclear radio jets are not common.

On kpc-scale, the fraction of radio luminous AGN is observed to be slightly higher. For example, among the sources catalogued in the 1.4 GHz VLA Faint Images of the Radio Sky at Twenty-Centimeters (FIRST) survey, about 10\% of optically selected AGN are found to be radio luminous and powered by relativistic jets in the radio \citep[e.g.,][]{White2000, Ivezic2002}. Also, \citet{Burlon2013} reported that $\sim$14\% of Seyfert-like sources were detected in ATCA 20 GHz blind survey among the ultra hard X-ray AGN sample. Considering the scale probed in this work using the VLBI, our lower detection of powerful AGN jets in pc-scale suggests that the nuclear jets are relatively shorter lived rather than the kpc-scale ones. This trend is consistent with the age measurements of jets in different physical scales; 10 $-$ 10$^{4}$ years for pc-scale nuclear jets and $\sim$10$^{6}$ years for kpc-scale extended jets \citep[e.g.,][]{Murgia2003, Venturi2004, Shabala2008, Nagai2009}.


\subsection{The relations of nuclear jet luminosity with the SMBH mass and accretion luminosity}

\begin{figure*}
	\includegraphics[width=\linewidth]{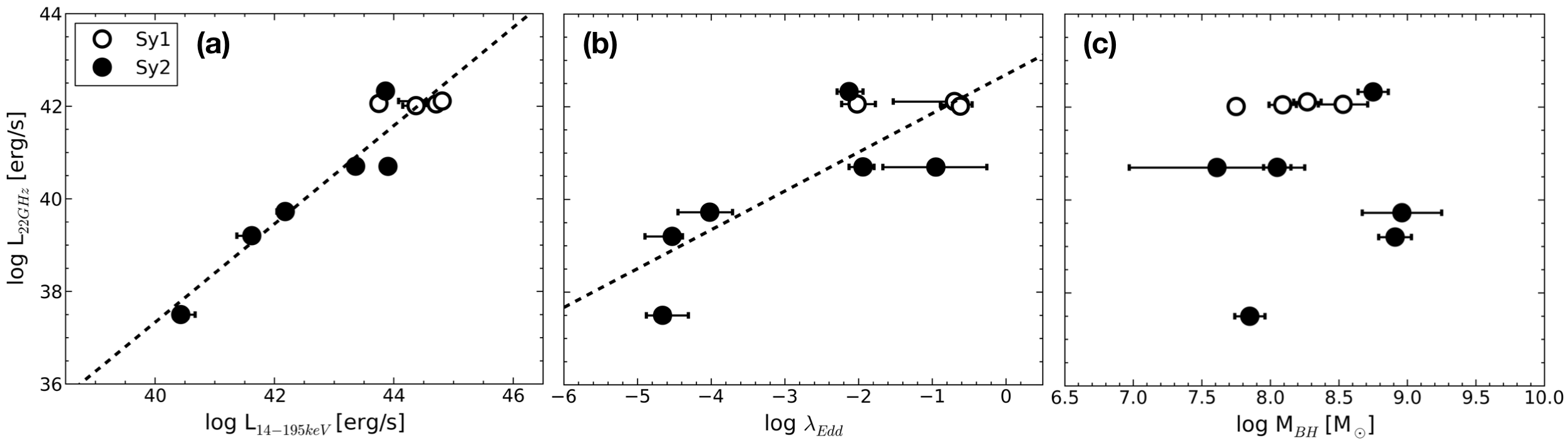}
    \caption{The relationships of VLBI 22 GHz luminosity with X-ray/BH properties of 10 targets. From left to right, (a) VLBI 22 GHz luminosity ($L_{\rm 22 GHz}$) vs. obscuration corrected ultra hard X-ray (14$-$195 keV) luminosity, (b) $L_{\rm 22 GHz}$ vs. Eddington ratio ($\equiv$ $L_{\rm bol}$/$L_{\rm Edd}$), and (c) $L_{\rm 22 GHz}$ vs. BH mass. Seyfert 1 and 2 types are indicated by open circles and filled circles, respectively. The correlation coefficient of each relation is (a) 0.95, (b) 0.84, and (c) -0.01, respectively. The black dashed line indicates the $\chi$-square best fit representing (a) log $L_{\rm 22GHz}$ = 1.06 log $L_{\rm 14-195keV}$ - 5.13 and (b) log $L_{\rm 22GHz}$ = 0.84 log $\lambda_{\rm Edd}$ + 42.70.}
    \label{fig:figure3}
\end{figure*}

\noindent In order to understand the powering mechanism of nuclear jets, we first probe the relationship between jet luminosity and accretion luminosity as well as BH mass. For this, it is important to have the BH mass which is accurate as possible. Depending on the method however, the discrepancy among the measurements from different studies on a single object can be quite large, several orders of magnitude in some extreme cases. To find the most reliable BH mass of our sample, we therefore revisited the literature including BASS DR1 \citep{Koss2017} and made a selection based on the following priorities: from highest to lowest, 1) gas or stellar kinematics of the central region or reverberation mapping, 2) $M_{\rm BH}-\sigma_{*}$ relation, and 3) broad H$\beta$ linewidth. For those cases where the same methodology was used in BASS DR1 and in the other studies, we adopted the BASS measurement. The final adopted BH masses and their measurement sources are listed in Table~\ref{tab:tab4}. For the X-ray luminosity, we adopted the obscuration corrected value at 14$-$150 keV from \citet{Ricci2017b} based on fitting 0.5$-$150 keV data. We then used a constant conversion of 1.14 to the more commonly used 14$-$195 keV value assuming $\Gamma \sim 1.8$.

The left panel of Figure~\ref{fig:figure3} shows the relationship between the obscuration corrected 14$-$195 keV luminosity and the 22 GHz radio luminosity of the pc-scale region of our AGN sample. These two quantities are correlated with the correlation coefficient of 0.95 and the Pearson test's p-value of 2.7$\times$10$^{-5}$. This tight correlation suggests that in the case where a (sub)pc-scale jet exists, its power is determined by the accretion luminosity and it also responses to the accretion activity in a relatively short timescale. This is also supported by the relation between the 22 GHz brightness and the accretion rate ($\lambda_{\rm Edd}$), i.e. the mass-normalized accretion luminosity as shown by the middle panel of Figure~\ref{fig:figure3}. However, $L_{\rm 22 GHz}$ in pc-scale is not correlated with the BH mass itself as seen in the right panel of Figure~\ref{fig:figure3}. 

It is worth noting that why previous VLBI studies did not find any clear relation between accretion luminosity and nuclear radio luminosity. For example, \citet{Panessa2013} measured the pc-scale radio luminosity of 28 nearby Seyfert galaxies ($z<0.005$) using the European VLBI Network (EVN) at 1.4 GHz and 5 GHz. But at these low frequencies, not only nonthermal synchrotron emission but also thermal free-free emission becomes more luminous with notably improved instrumental sensitivity compared to in high radio frequencies. Therefore their measurements could have well been contaminated by the emission from corona and winds. Indeed, the broad range of brightness temperature and $L_{R}$/$L_{X}$ of their sample (10$^{5}$ $-$ 10$^{10}$ K, and 10$^{-6}$ $-$ 10$^{-3}$, respectively) are highly suggestive of the various origins of nuclear radio emission.

\subsection{Fundamental plane of black hole activity}

\begin{figure}
	\includegraphics[width=\columnwidth]{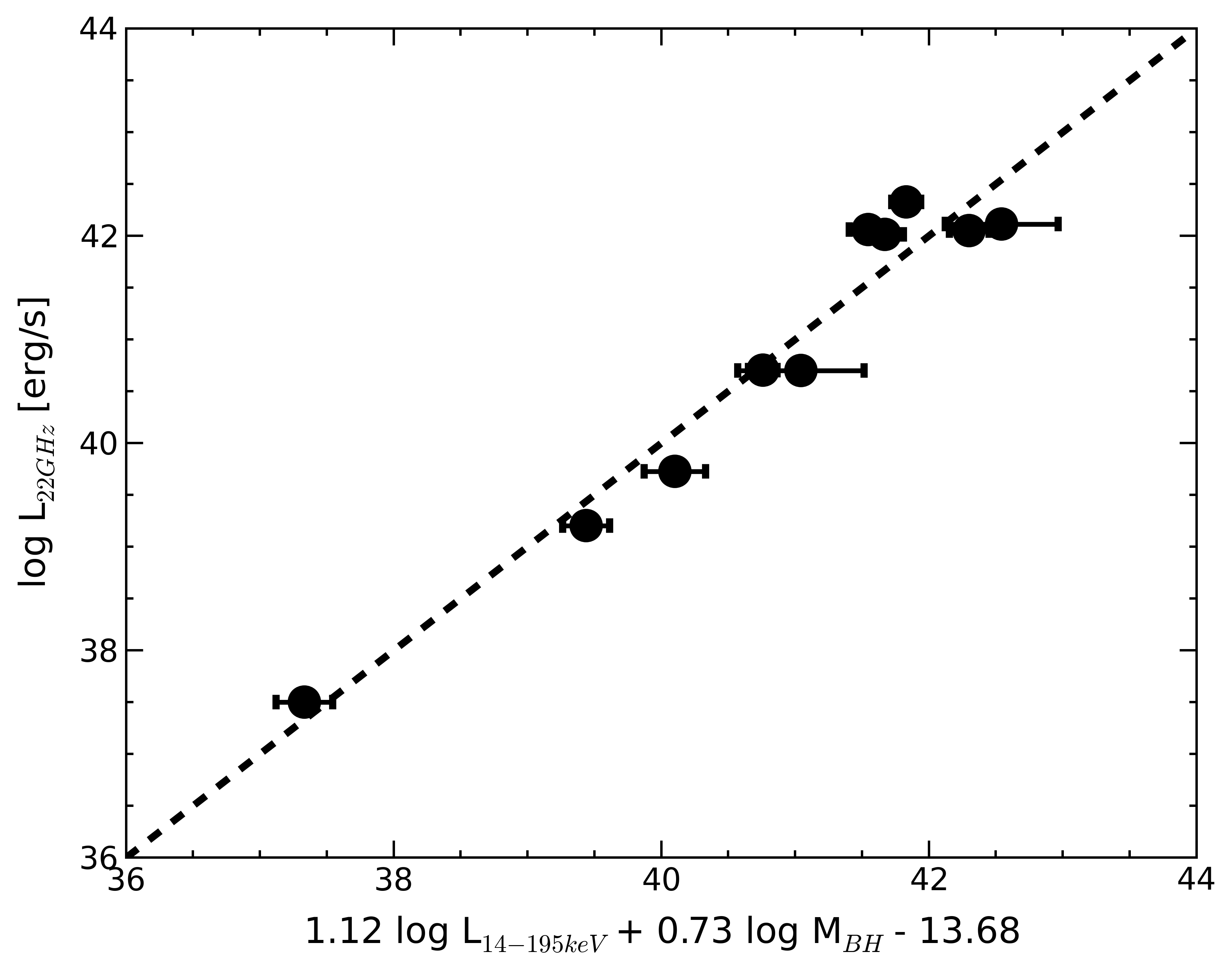}
    \caption{A fundamental plane of black hole activities in pc-scale. Our 10 VLBI targets are nicely found in a single plane (log $L_{\rm 22GHz,VLBI}$ = 1.12 log $L_{\rm 14-195 keV}$ + 0.73 log $M_{\rm BH}$ - 13.68) in the space of VLBI $L_{\rm 22GHz}$, $L_{\rm 14-195 keV}$ and $M_{\rm BH}$. Compared to the canonical fundamental plane with $\sigma_{\rm R}$ = 0.88 \citep{Merloni2003}, our VLBI data-based fundamental plane shows a smaller scatter of $\sigma_{\rm VLBI}$ = 0.35.}
    \label{fig:figure4}
\end{figure}

\noindent It is interesting that the radio luminosity is more significantly correlated with accretion luminosity and BH mass. For example, \citet{Merloni2003} probed 2$-$10 keV luminosity, BH mass and 5 GHz luminosity (a proxy of kpc-scale jet luminosity) of various AGN populations, finding a good linear correlation among three quantities which not only SMBHs but also stellar mass BHs follow. This scaling relation (log $L_{\rm 5 GHz}$ = 0.60 log $L_{\rm 2-10 keV}$ + 0.78 log $M_{\rm BH}$ + 7.33), which is known as a fundamental plane of black hole activity, implies an important connection of the synchrotron jet with the other two observables.

Surprisingly, this correlation seems to be valid in the very vicinity of SMBHs down to pc-scale. As shown in Figure~\ref{fig:figure4}, we find a tight correlation among the VLBI 22 GHz luminosity, the ultra hard X-ray luminosity at 14$-$195 keV and the BH mass, with the multiple linear regression fit as follows:

\begin{flushleft}
log $L_{\rm 22GHz,VLBI}$ = (1.12 $\pm$ 0.10) log $L_{\rm 14-195 keV}$
\end{flushleft}
\vspace{-4mm}
\begin{flushright}
+ (0.73 $\pm$ 0.29) log $M_{\rm BH}$ - (13.68 $\pm$ 5.42)
\end{flushright}

The tight correlation with pc-scale synchrotron radiation ($\xi_{\rm RX}$=1.12, $\xi_{\rm RM}$=0.73, $\sigma_{\rm VLBI}$=0.35) compared to the scaling relation in kpc-scale of \citet{Merloni2003} ($\xi_{\rm RX}$=0.60, $\xi_{\rm RM}$=0.78, $\sigma_{\rm R}$=0.88) may indicate that the power of nuclear jet is more relevant to the accretion activity. As shown in Figure~\ref{fig:figure3} (c), $L_{\rm 22GHz}$ is not correlated with $M_{\rm BH}$. However, the strong correlation between $L_{\rm 22GHz}$ and $L_{14-195keV}$ yields a tight relationship among the three in the fundamental plane as seen in Figure~\ref{fig:figure4}. From this, we can infer that the accretion history is episodic and the current accretion activity is not highly dependent of the past accretion event(s), if the current SMBH mass is the result of the accretion history in the past. On the other hand, our sample size is small, and does not include a broad mass range of BHs with $M_{\rm BH}$ ranging only from 10$^{7}$ to 10$^{9}$ $M_{\odot}$ unlike the sample of \citet{Merloni2003} which spans from 10$^{2}$ to 10$^{10}$ $M_{\odot}$. The narrow range of BH mass among our sample could also result in the smaller scatter of the VLBI data-based fundamental plane.

\section{DISCUSSION}

\subsection{Evolution sequence of AGN}

\begin{figure}
	\includegraphics[width=\columnwidth]{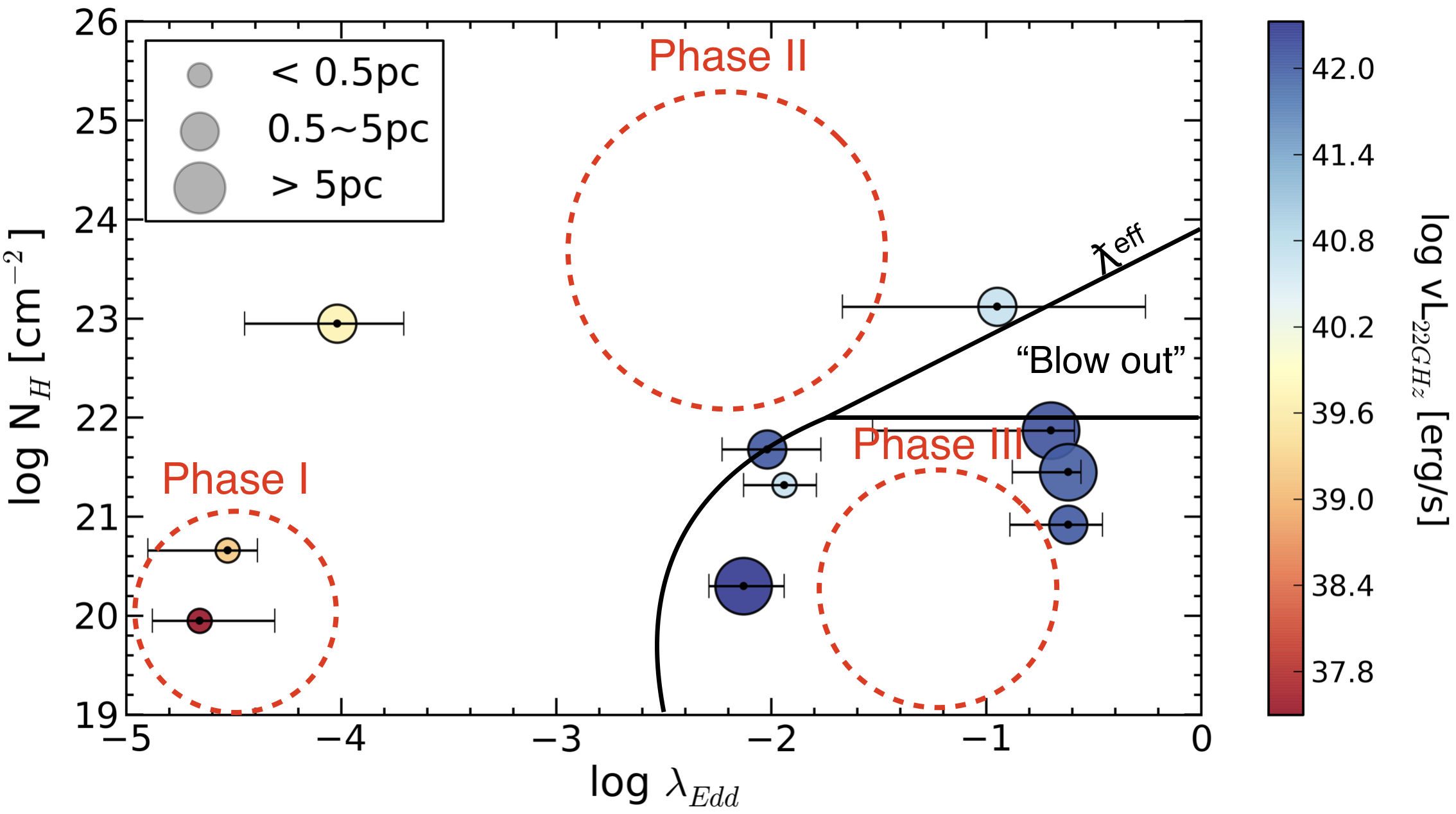}
    \caption{Our sample distribution in $N_{\rm H}$-$\lambda_{\rm Edd}$ space. Colours represent VLBI 22 GHz luminosities with red for being less luminous and blue for being more luminous. The size of circles indicates the physical extent of the jet in pc-scale. Phase \rom{1}, \rom{2} and \rom{3} of the AGN evolution sequence proposed by \citet{Ricci2017a} are indicated by red-dashed circles. Both the luminosity and the physical extent of the pc-scale jet of our sample gradually increase from Phase \rom{1} to \rom{2} then, \rom{3}.}
    \label{fig:figure5}
\end{figure}

\noindent Once the matter starts to be accreted onto the BH, both the accretion rate and the gas column density will increase until the radiation pressure of the AGN balances with the anchoring pressure of ambient dusty gas, that is, the effective Eddington limit, $\lambda_{\rm Edd}^{\rm eff}$ is reached. At this stage, further accretion could expel most of the obscuring material and hence the AGN rapidly becomes unobscured, forming a zone of avoidance where both the density around the BH and the accretion rate are high. This evolutionary sequence was observed by \citet{Ricci2017a} in the relation between hydrogen column density and Eddington rate among the X-ray selected AGN \citep[see Figure 3 in][]{Ricci2017a} based on earlier theoretical work by \citet{Fabian2008, Fabian2009}.

In Figure~\ref{fig:figure5}, our targets are shown on the top of $N_{\rm H}$ - $\lambda_{\rm Edd}$ relation from \citet{Ricci2017a}, with $N_{\rm H}$ from \citet{Ricci2017b}. Noticeably, none of the sources of our sample is found in the ``Blow-out" region beyond $\lambda_{\rm Edd}^{\rm eff}$, i.e. a zone of avoidance, likely due to the AGN feedback. It becomes more interesting when our sample is probed by their VLBI luminosity, which we roughly divided into three groups. The ones with low luminosity (red and yellow) are found somewhere between Phase \rom{1} and Phase \rom{2} with the least luminous VLBI source in the corner of Phase \rom{1}. Whereas, relatively strong VLBI targets are mostly found in Phase \rom{3}, which is thought to be after the ``Blow-out" phase. 

The location of our sample in the $N_{\rm H}$ - $\lambda_{\rm Edd}$ space roughly follows what is predicted by the evolutionary sequence based on \citet{Ricci2017a} in a sense that the sample moves from Phase \rom{1} to Phase \rom{2}, then to Phase \rom{3} as the nuclear radio luminosity increases. In addition, the measured physical size of nuclear jets in our pc-scale VLBI image also gradually increases along the evolutionary sequence. In particular, among our sample three Fanaroff-Riley type AGN \citep[FR;][]{Fanaroff1974} with a very extended kpc-scale jet, i.e. aged AGN (3C 111.0, 3C 120, and NGC 1275) are all found in Phase \rom{3} of Figure~\ref{fig:figure5}. So the distribution of our sample in the $N_{\rm H}$ - $\lambda_{\rm Edd}$ space, and the location of FR type AGN support the SMBH evolution sequence suggested by \citet{Ricci2017a}, even though the size of the sample is rather small.

\subsection{Jet triggering source}

\begin{figure}
	\includegraphics[width=\columnwidth]{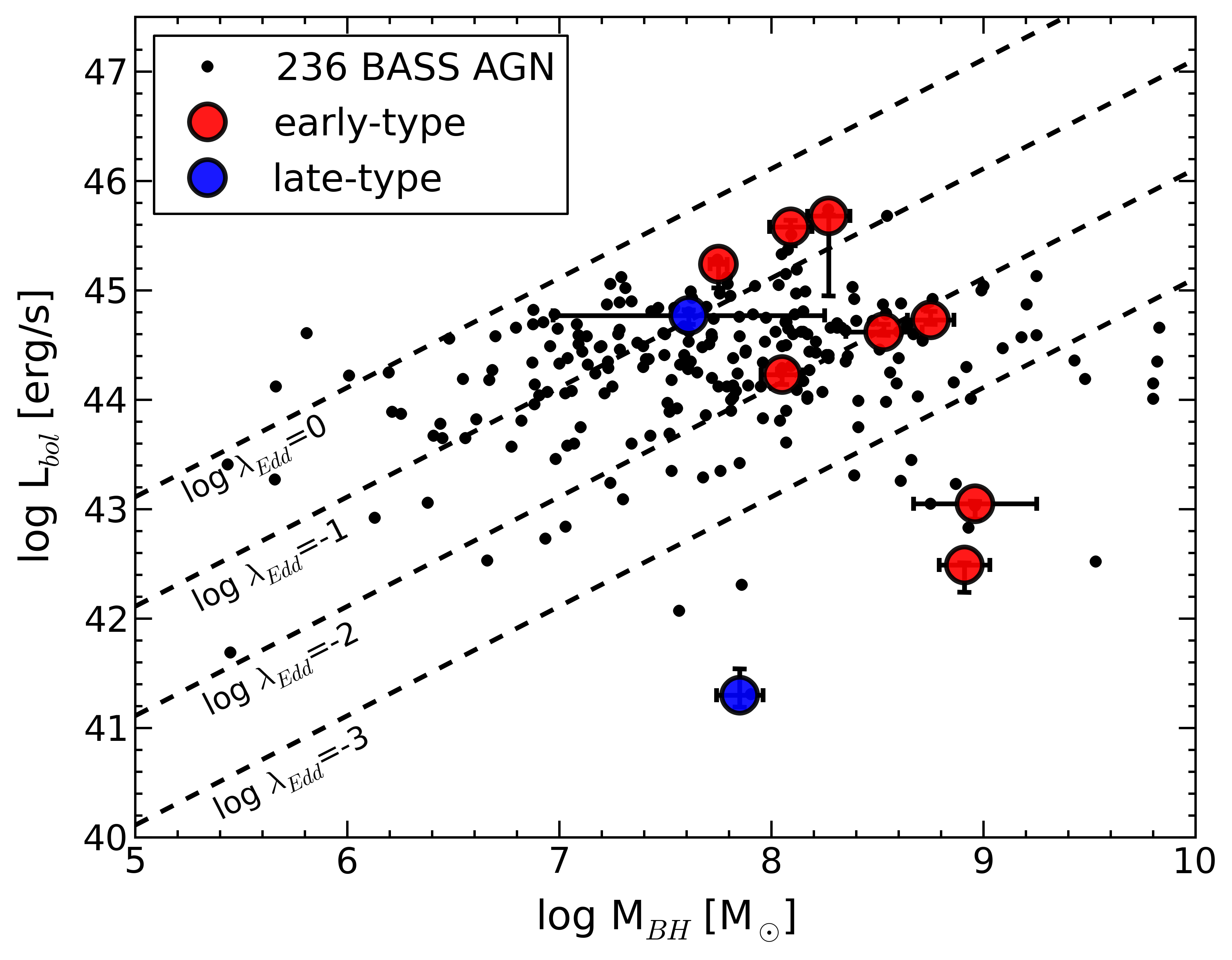}
    \caption{The distribution of our sample in the $L_{\rm bol}$-$M_{\rm BH}$ space. Small black dots represent 236 BASS AGN and large red/blue dots indicate early/late-type of our sample. Our radio-active, but hosted by galaxies of various properties, are broadly span in this relation.}
    \label{fig:figure7}
\end{figure}

\noindent In this section, we investigate the potential triggering mechanism for the powerful nuclear jet activity. Figure~\ref{fig:figure7} shows the relationship of $L_{\rm bol}$ and $M_{\rm BH}$ of our VLBI sample. Compared to the BASS parent sample, our targets appear to host relatively massive SMBHs on average ($M_{\rm BH}$ > 10$^{7.5}$ $M_{\odot}$). It may imply that the mass of a BH is an important condition to launch a nuclear jet. However, there is no linear correlation between these two quantities (see Figure~\ref{fig:figure3}(c)), and our sample shows a broad range of X-ray luminosity and Eddington rate. This may indicate that the amount of mass accretion rate, which can be a tracer of the accretion disc geometry, is not the only critical requirement in triggering a nuclear jet, and the jet can be powered by various origins. Indeed, the host galaxy properties and the environment of our sample are quire intriguing as described below.

- \textbf{NGC 1275} is a cD galaxy located in the center of the Perseus cluster. The surrounding H$\alpha$ filaments are likely to have formed through the compression of the intra cluster gas. The outer part of the stellar disc is very blue, implying the presence of young star clusters, which could have formed from the recent accretion of a gas-rich system \citep{Conselice2001}.

- \textbf{3C 120} is a Seyfert 1 type hosted by an edge-on S0 galaxy. It reveals a very extended H$\alpha$ filament of $\sim$60 kpc \citep{Heckman1979}, which might be an indication of gas inflow. Its jets are known to be interacting with surrounding clouds, which has been suggested to be the result of the collision between galaxies \citep{Gomez2006}.

- \textbf{Mrk 348} is a huge spiral galaxy with the evidence for tidal disruption as revealed by its H{\sc i} velocity field. Stellar plumes which are likely to have a tidal origin have been also detected in the optical \citep{Simkin1987, Baum1993}.

- \textbf{NGC 1052} is an elliptical galaxy with an exceptionally large H{\sc i} extent, which is about 3 times larger than its optical size ($D_{\rm H\rom{1}}$ = 16 kpc). Both the morphological and the kinematical properties of H{\sc i} gas are highly suggestive of its capturing the gas from gas-rich dwarf or spiral neighbor \citep{vanGorkom1986}. The misalignment of pc-scale and kpc-scale radio jet has been reported \citep{Vermeulen2003,Kadler2004}.

- \textbf{M 81} is found in relatively busy environments with a number of close neighbouring galaxies. The gas exchange with neighbors has been confirmed by H{\sc i} imaging data, indicating that they are in the early to advanced merging state \citep[e.g.,][]{Yun1994}.

- \textbf{NGC 3998} is one of the H{\sc i}-richest systems among 166 early-type galaxies studied in a volume-limited H{\sc i} survey, ATLAS 3D \citep{Cappellari2011}. Its H{\sc i} morphology is highly disturbed \citep{Serra2012}, and a polar H{\sc i}/H$\alpha$ warp in the core region has been seen \citep{Frank2016}, indicating tidal interactions.

- \textbf{Arp 102B} is an E0 which is classified as LINER/Seyfert 1. A two-fold mini spiral structure is found in H$\alpha$ in the inner $\sim$1 kpc, which could have been formed by a recent gas accretion event \citep{Fathi2011}. This also might have triggered some nuclear activity and formed a jet which seems to be pushing the circumnuclear H$\alpha$ gas \citep{Couto2013}.

- \textbf{PKS 2331-240} is a giant radio galaxy. Its AGN activity is likely to have been recently repowered \citep{Hernandez-Garcia2017}.

No detailed studies can be found on the remaining two targets (Q0241+622 and 3C 111.0), yet it is worth to be noted that the majority of our sample are found with some hint or direct evidence for gas inflow from surroundings. This may imply that the external process of host galaxy can be an important mechanism to launch the synchrotron jet on SMBHs.\\

In the BASS collaboration, some parallel studies on the radio properties of BASS AGN are currently being carried on. In this work, we probed the 22 GHz parsec-scale nuclear jet properties of 10 radio-luminous AGN, whereas Smith et al. studied 22 GHz kpc-scale radio properties of 96 star-formation/radio-quiet AGN (submitted). Meanwhile, B\"{a}r et al. investigated the characteristics of the most X-ray luminous 28 Seyfert 2 type AGN in multi-wavelengths including 1.4 GHz (submitted). In spite of the common aspects shared by these works (i.e. the same parent BASS sample in low-z and the employment of radio data), individual studies probe fairly different sub-population of X-ray selected AGN. We expect all these efforts to help us understanding the radio properties of AGN in the broader context in the future.

\section{SUMMARY}

\noindent In this work, we have presented the results from our VLBI 22 GHz imaging study of X-ray/radio selected AGN in the local Universe (z < 0.05). The main goal of this study is to investigate the correlation between accretion activity and jet properties in the vicinity of SMBH in (sub)pc-scale. In particular, we probed the strength and the morphology of pc-scale radio emission together with the obscuration corrected ultra hard X-ray luminosity \citep{Ricci2017b} and BH mass \citep{Koss2017}. Our results are:

\begin{itemize}
\item About 4\% of X-ray selected AGN in the local Universe (z < 0.05) are found with a pc-scale extended feature at 22 GHz. This implies that a powerful nuclear jet is not common among local AGN.\\

\item The parsec-scale morphology of our sample suggests that their radio emission mainly comes from the synchrotron jets rather than the thermal radiation or coronal emission.\\

\item Once the nuclear jet is formed, the nuclear jet power is like to be determined by the radiation from the mass accretion, within a short timescale.\\

\item We find a tight correlation among our sample in the fundamenatl plane  (pc-scale radio luminosity, X-ray luminosity, and BH mass space) of black hole activity. This suggests that launching of nuclear jet is highly correlated with accretion activity.\\

\item In $N_{\rm H}$ and $\lambda_{\rm Edd}$ space, both jet power and size in parsec-scale are observed to increase from low $N_{\rm H}$ and $\lambda_{\rm Edd}$, to higher $N_{\rm H}$ and $\lambda_{\rm Edd}$, then low $N_{\rm H}$ and mid/high $\lambda_{\rm Edd}$, supporting the AGN evolutionary sequence proposed by previous studies \citep{Fabian2008,Fabian2009,Ricci2017a}.\\

\item A majority of the host galaxies of our AGN are found with some hints of gas inflow or galaxy-galaxy merging. This suggests that the external interaction of galaxy scale can be an important launching mechanism of the synchrotron jet of SMBHs.\\

\end{itemize}

\section*{Acknowledgements}

We thank the anonymous referee for his/her valuable comments and suggestions that helped to improve the paper. We are grateful to the staff of the KVN who helped to operate the array and to correlate the data. The KVN and a high-performance computing cluster are facilities operated by the KASI (Korea Astronomy and Space Science Institute). The KVN observations and correlations are supported through the high-speed network connections among the KVN sites provided by the KREONET (Korea Research Environment Open NETwork), which is managed and operated by the KISTI (Korea Institute of Science and Technology Information). In addition, this work is based on observations made with the KaVA and VLBA, which are operated by the the Korea Astronomy and Space Science Institute and the National Astronomical Observatory of Japan and the National Radio Astronomy Observatory which is a facility of the National Science Foundation operated under cooperative agreement by Associated Universities, Inc., respectively. The VLBA data was correlated by the Swinburne University of Technology software correlator, developed as part of the Australian Major National Research Facilities Programme and operated under licence. Support for this work was provided by the National Research Foundation of Korea to the Center for Galaxy Evolution Research (No. 2017R1A5A1070354), and NRF grant No. 2015R1D1A1A0106051 and 2018R1D1A1B07048314. J.B. acknowledges support from NRF through Young Researchers' Exchange Program Between Korea and Switzerland 2016 (No. 2016K1A3A1A14953055). K.O. acknowledges support from the Japan Society for the Promotion of Science (JSPS, ID: 17321). M.K. acknowledges support from NASA through ADAP award NNH16CT03C. C.R. acknowledges the CONICYT+PAI Convocatoria Nacional subvencion a instalacion en la academia convocatoria a\~{n}o 2017 PAI77170080. Y.U. acknowledges the financial support by JSPS Grant-in-Aid for Scientific Research 17K05384.




\bibliographystyle{mnras}




\appendix

\setcounter{table}{0}
\renewcommand{\thetable}{A\arabic{table}}

\section*{APPENDIX}

Further details of our 22 GHz KVN fringe survey and the archival VLBA data of 47 GB6/PMN cross-matched AGN can be found in Table~\ref{tab:tabA1}. The full list of 95 targets for the 15 GHz VLBA blind fringe survey is provided in Table~\ref{tab:tabA2}.

\begin{table*}
	\centering
	\caption{List of 22 GHz fringe surveys for 47 GB6/PMN cross-matched sources}
	\label{tab:tabA1}
	\resizebox{0.9\textwidth}{!}{
	\begin{tabular}{cccccccc} 
		\hline
		\hline
		BAT & Name & R.A. & Decl. & redshift & $S_{\rm 5GHz}$ & Obs.code & Obs.date\\
		index & & (hh:mm:ss.sss) & (dd:mm:ss.ss) & & (mJy) & (VLBI) & (yyyy.mm.dd) \\
		(1) & (2) & (3) & (4) & (5) & (6) & (7) & (8)\\
		\hline
		\hline
		\multicolumn{8}{l}{\textbf{\emph{22 GHz VLBI detections}}}\\
		33 & Mrk 348 & 00:48:47.142 & +31:57:25.08 & 0.0150 & 302 $\pm$ 27 & VLBA/BP182 & 2014.04.20.\\
		140 & NGC 1052 & 02:41:04.799 & $-$08:15:20.75 & 0.0045 & 3158 $\pm$ 99 & VLBA/BR130H & 2009.03.08.\\
		147 & Q0241+622 & 02:44:57.697 & +62:28:06.52 & 0.0435 & 376 $\pm$ 33 & VLBA/BH136F & 2006.11.29.\\
		173 & NGC 1275 & 03:19:48.160 & +41:30:42.11 & 0.0166 & 46894 $\pm$ 4179 & VLBA/BA105 & 2013.02.16.\\
		214 & 3C 111.0 & 04:18:21.277 & +38:01:35.80 & 0.0500 & 5168 $\pm$ 460 & VLBA/BT104 & 2009.12.18.\\ 
		226 & 3C 120 & 04:33:11.096 & +05:21:15.62 & 0.0330 & 5189 $\pm$ 99 & VLBA/BG182B & 2007.11.07.\\
		477 & M 81 & 09:55:33.173 & +69:03:55.06 & -0.0001 & 98 $\pm$ 10 & VLBA/BH173 & 2001.10.03.\\
		579 & NGC 3998 & 11:57:56.133 & +55:27:12.93 & 0.0034 & 88 $\pm$ 8 & KVN/N16JB02A & 2016.03.24.\\ 
		876 & ARP 102B & 17:19:14.490 & +48:58:49.44 & 0.0245 & 164 $\pm$ 15 & KVN/N16JB02A & 2016.03.24.\\
		1200 & PKS 2331-240 & 23:33:55.238 & $-$23:43:40.66 & 0.0477 & 907 $\pm$ 48 & VLBA/BS229 & 2014.08.15.\\
        \hline
		\multicolumn{8}{l}{\textbf{\emph{22 GHz VLBI non-detections}}}\\
		28 & NGC 235A & 00:42:52.810 & $-$23:32:27.71 & 0.0221 & 56 $\pm$ 11 & KVN/N16JB02A & 2016.03.24.\\
		74 & NGC 513 & 01:24:26.806 & +33:47:58.24 & 0.0191 & 22 $\pm$ 4 & KVN/N16JB02A & 2016.03.24.\\
		144 & NGC 1068 & 02:42:40.771 & $-$00:00:47.84 & 0.0030 & 2039 $\pm$ 99 & VLBA/GG042 & 2000.04.24.\\ 
		308 & NGC 2110 & 05:52:11.376 & $-$07:27:22.49 & 0.0074 & 165 $\pm$ 14 & KVN/N16JB02A & 2016.03.24.\\
		310 & UGC 3374 & 05:54:53.609 & +46:26:21.63 & 0.0202 & 83 $\pm$ 8 & KVN/N16JB02A & 2016.03.24.\\ 
		317 & UGC 3386 & 06:02:37.988 & +65:22:16.46 & 0.0175 & 46 $\pm$ 5 & KVN/N16JB02A & 2016.03.24.\\
		325 & Mrk 3 & 06:15:36.458 & +71:02:15.24 & 0.0134 & 363 $\pm$ 32 & KVN/N16JB02A & 2016.03.24.\\
		347 & Mrk 6 & 06:52:12.323 & +74:25:37.24 & 0.0190 & 105 $\pm$ 10 & KVN/N16JB02A & 2016.03.24.\\ 
		348 & 2MASS J0654$^{*}$ & 06:54:34.186 & +07:03:20.94 & 0.0240 & 37 $\pm$ 6 & KVN/N16JB02A & 2016.03.24.\\
		404 & Mrk 1210 & 08:04:05.862 & +05:06:49.81 & 0.0135 & 76 $\pm$ 11 & VLBA/BK163B & 2010.09.06.\\
		439 & Mrk 18 & 09:01:58.405 & +60:09:06.23 & 0.0110 & 25 $\pm$ 4 & KVN/N16JB02A & 2016.03.24.\\ 
		467 & UGC 5101 & 09:35:51.694 & +61:21:10.52 & 0.0393 & 77 $\pm$ 7 & KVN/N16JB02A & 2016.03.24.\\ 
		471 & NGC 2992 & 09:45:42.045 & $-$14:19:34.90 & 0.0076 & 102 $\pm$ 12 & KVN/N16JB02A & 2016.03.24.\\
		480 & NGC 3081 & 09:59:29.544 & $-$22:49:34.75 & 0.0080 & 45 $\pm$ 11 & KVN/N16JB02A & 2016.03.24.\\
		484 & NGC 3079 & 10:01:57.803 & +55:40:47.24 & 0.0034 & 320 $\pm$ 28 & VLBA/BM208D & 2005.08.18.\\ 
		497 & NGC 3227 & 10:23:30.570 & +19:51:54.30 & 0.0033 & 48 $\pm$ 6 & KVN/N16JB02A & 2016.03.24.\\
		518 & NGC 3393 & 10:48:23.467 & $-$25:09:43.30 & 0.0129 & 52 $\pm$ 11 & VLBA/BG169 & 2007.01.31.\\
		524 & Mrk 728 & 11:01:01.774 & +11:02:48.91 & 0.0356 & 45 $\pm$ 6 & KVN/N16JB02A & 2016.03.24.\\
		548 & NGC 3718 & 11:32:34.853 & +53:04:04:49 & 0.0028 & 19 $\pm$ 4 & KVN/N16JB02A & 2016.03.24.\\
		560 & NGC 3786 & 11:39:42.514 & +31:54:33.96 & 0.0090 & 20 $\pm$ 4 & KVN/N16JB02A & 2016.03.24.\\
		585 & NGC 4051 & 12:03:09.610 & +44:31:52.69 & 0.0020 & 22 $\pm$ 4 & VLBA/BT117A & 2011.07.16.\\
		590 & NGC 4102 & 12:06:23.115 & +52:42:39.42 & 0.0018 & 85 $\pm$ 8 & KVN/N16JB02A & 2016.03.24.\\
		595 & NGC 4151 & 12:10:32.577 & +39:24:21.06 & 0.0031 & 139 $\pm$ 13 & KVN/N16JB02A & 2016.03.24.\\ 
		609 & NGC 4258 & 12:18:57.620 & +47:18:13.39 & 0.0017 & 174 $\pm$ 16 & VLBA/TY027 & 2013.01.30.\\ 
		615 & NGC 4388 & 12:25:46.820 & +12:39:43.45 & 0.0083 & 85 $\pm$ 10 & VLBA/BB258D & 2008.09.22.\\
		665 & NGC 5033 & 13:13:27.535 & +36:35:37.14 & 0.0027 & 79 $\pm$ 8 & KVN/N16JB02A & 2016.03.24.\\
		669 & NGC 5106 & 13:20:59.612 & +08:58:42.16 & 0.0326 & 50 $\pm$ 8 & KVN/N16JB02A & 2016.03.24.\\ 
		670 & PGC 46710 & 13:22:24.485 & $-$16:43:42.09 & 0.0168 & 104 $\pm$ 12 & KVN/N16JB02A & 2016.03.24.\\
		738 & Mrk 477 & 14:40:38.098 & +53:30:16.24 & 0.0377 & 21 $\pm$ 4 & KVN/N16JB02A & 2016.03.24.\\
		772 & PGC 55410 & 15:33:20.698 & $-$08:42:01.77 & 0.0228 & 103 $\pm$ 12 & KVN/N16JB02A & 2016.03.24.\\ 
		836 & LEDA 214543 & 16:50:42.752 & +04:36:18.30 & 0.0319 & 34 $\pm$ 7 & KVN/N16JB02A & 2016.03.24.\\
		841 & NGC 6240 & 16:52:58.861 & +02:24:03.55 & 0.0239 & 164 $\pm$ 15 & VLBA/BH088 & 2001.10.03.\\ 
		960 & PGC 61662 & 18:16:11.627 & +42:39:37.25 & 0.0409 & 24 $\pm$ 4 & KVN/N16JB02A & 2016.03.24.\\
		1077 & PGC 64775 & 20:28:35.061 & +25:44:00.18 & 0.0139 & 30 $\pm$ 5 & KVN/N16JB02A & 2016.03.24.\\
		1158 & NGC 7319 & 22:36:03.602 & +33:58:33.18 & 0.0225 & 27 $\pm$ 4 & KVN/N16JB02A & 2016.03.24.\\
		1182 & NGC 7469 & 23:03:15.674 & +08:52:25.28 & 0.0160 & 95 $\pm$ 12 & KVN/N16JB02A & 2016.03.24.\\
		1184 & NGC 7479 & 23:04:56.668 & +12:19:22.36 & 0.0071 & 41 $\pm$ 6 & KVN/N16JB02A & 2016.03.24.\\
		\hline
		\hline
	\end{tabular}
	}
	\begin{flushleft}
	Note. (1) Entry number listed in the {\it Swift} BAT 70-month hard X-ray survey \citep{Baumgartner2013}; (2) Source name; (3), (4), and (5) Right ascension and declination in J2000 coordinate, and redshift obtained from the SIMBAD Astronomical Database; (6) 5 GHz flux density and error from GB6 \citep{Gregory1996} and PMN \citep{Griffith1993} catalogs; (7) and (8) Observation code and date of our 22 GHz KVN fringe survey and archival VLBA data.\\
	$^{*}$ 2MASS J06543417+0703210
	\end{flushleft}
\end{table*}

\begin{table*}
	\centering
	\caption{List of 95 targets in our VLBA blind fringe survey at 15 GHz}
	\label{tab:tabA2}
	\resizebox{\textwidth}{!}{
	\begin{tabular}{ccccc|ccccc} 
		\hline
		\hline
		BAT & Name & R.A. & Decl. & redshift & BAT & Name & R.A. & Decl. & redshift \\
		index & & (hh:mm:ss.sss) & (dd:mm:ss.ss) & & index & & (hh:mm:ss.sss) & (dd:mm:ss.ss) & \\
		(1) & (2) & (3) & (4) & (5) & (1) & (2) & (3) & (4) & (5) \\
		\hline
		\hline
        1 & LEDA 1023662 & 00:00:48.765 & $-$07:09:11.75 & 0.0375 & 503 & LEDA 31154 & 10:32:44.513 & $-$28:36:35.93 & 0.0121 \\
        4 & LEDA 1814347 & 00:03:27.428 & +27:39:17.38 & 0.0397 & 504 & LEDA 31274 & 10:34:23.668 & +73:00:49.91 & 0.0224 \\
        6 & Mrk 335 & 00:06:19.582 & +20:12:10.58 & 0.0259 & 512 & SWIFT J1043.4+1105 & 10:43:26.466 & +11:05:24.24 & 0.0480 \\
        13 & LEDA 136991 & 00:25:32.928 & +68:21:44.30 & 0.0125 & 513 & LEDA 31994 & 10:44:08.489 & +70:24:19.43 & 0.0320 \\
        24 & Mrk 334 & 00:38:32.149 & +23:36:47.55 & 0.0254 & 515 & LEDA 32071 & 10:44:48.999 & +38:10:52.53 & 0.0261 \\
        31 & LEDA 2573 & 00:43:08.757 & $-$11:36:03.35 & 0.0192 & 517 & LEDA 32188 & 10:46:42.477 & +25:55:54.00 & 0.0205 \\
        43 & Mrk 352 & 00:59:53.309 & +31:49:37.17 & 0.0152 & 519 & Mrk 417 & 10:49:30.890 & +22:57:52.37 & 0.0326 \\
        53 & LEDA 3938 & 01:06:45.236 & +06:38:01.59 & 0.0410 & 520 & NGC 3431 & 10:51:15.037 & $-$17:00:28.65 & 0.0174 \\
        55 & LEDA 963299 & 01:07:39.635 & $-$11:39:11.76 & 0.0475 & 528 & LEDA 33568 & 11:05:58.978 & +58:56:45.65 & 0.0477 \\
        60 & Mrk 975 & 01:13:51.000 & +13:16:18.63 & 0.0489 & 530 & NGC 3516 & 11:06:47.494 & +72:34:06.70 & 0.0087 \\
        64 & NGC 452 & 01:16:14.850 & +31:02:01.92 & 0.0177 & 531 & LEDA 33949 & 11:10:47.974 & $-$28:30:03.91 & 0.0238 \\
        70 & LEDA 5003 & 01:22:34.427 & +50:03:18.01 & 0.0206 & 532 & Mrk 732 & 11:13:49.721 & +09:35:10.68 & 0.0293 \\
        77 & Mrk 359 & 01:27:32.552 & +19:10:43.74 & 0.0167 & 533 & SWIFT J1114.3+2020 & 11:14:02.454 & +20:23:14.09 & 0.0262 \\
        79 & LEDA 5486 & 01:28:24.467 & +16:27:33.43 & 0.0387 & 534 & LEDA 34261 & 11:14:43.857 & +79:43:35.73 & 0.0372 \\
        96 & LEDA 6966 & 01:52:49.004 & $-$03:26:48.56 & 0.0172 & 542 & Mrk 40 & 11:25:36.150 & +54:22:57.31 & 0.0209 \\
        99 & LEDA 2295246 & 01:57:10.952 & +47:15:59.16 & 0.0478 & 543 & Mrk 423 & 11:26:48.523 & +35:15:02.98 & 0.0322 \\
        102 & NGC 788 & 02:01:06.450 & $-$06:48:56.98 & 0.0137 & 549 & IC 2921 & 11:32:49.288 & +10:17:47.39 & 0.0440 \\
        106 & Mrk 1018 & 02:06:16.006 & $-$00:17:29.23 & 0.0430 & 552 & Mrk 739E & 11:36:29.300 & +21:35:45.60 & 0.0297 \\
        113 & LEDA 138501 & 02:09:37.402 & +52:26:39.64 & 0.0494 & 554 & LEDA 1735060 & 11:38:33.706 & +25:23:53.20 & 0.0254 \\
        116 & Mrk 590 & 02:14:33.579 & $-$00:46:00.28 & 0.0266 & 557 & LEDA 801745 & 11:38:51.042 & $-$23:21:35.35 & 0.0272 \\
        119 & SWIFT J0216.3+5128 & 02:16:29.879 & +51:26:24.69 & 0.0288 & 565 & LEDA 36541 & 11:44:29.874 & +36:53:08.64 & 0.0386 \\
        129 & NGC 931 & 02:28:14.462 & +31:18:41.44 & 0.0163 & 566 & LEDA 36651 & 11:45:16.007 & +79:40:53.41 & 0.0063 \\
        130 & Mrk 1044 & 02:30:05.543 & $-$08:59:53.55 & 0.0159 & 567 & LEDA 88639 & 11:45:40.455 & $-$18:27:14.98 & 0.0326 \\
        133 & NGC 973 & 02:34:20.101 & +32:30:20.01 & 0.0160 & 568 & LEDA 36655 & 11:45:33.175 & +58:58:40.89 & 0.0100 \\
        134 & NGC 985 & 02:34:37.882 & $-$08:47:17.02 & 0.0430 & 576 & LEDA 37161 & 11:52:03.545 & $-$11:22:24.21 & 0.0500 \\
        145 & LEDA 2442097 & 02:44:02.962 & +53:28:28.17 & 0.0364 & 580 & LEDA 37721 & 11:58:52.555 & +42:34:13.66 & 0.0314 \\
        416 & LEDA 23515 & 08:23:01.100 & $-$04:56:05.39 & 0.0222 & 582 & LEDA 37894 & 12:00:57.924 & +06:48:22.69 & 0.0359 \\
        434 & LEDA 25044 & 08:55:12.572 & +64:23:45.17 & 0.0362 & 583 & Mrk 1310 & 12:01:14.348 & $-$03:40:41.08 & 0.0198 \\
        436 & NGC 2655 & 08:55:37.731 & +78:13:23.10 & 0.0047 & 1110 & 4C 50.55 & 21:24:39.400 & +50:58:25.00 & 0.0151 \\
        437 & NGC 2712 & 08:59:30.458 & +44:54:50.40 & 0.0060 & 1117 & SWIFT J2156.1+4728 & 21:35:53.994 & +47:28:21.74 & 0.0253 \\
        443 & SWIFT J0904.3+5538 & 09:04:36.971 & +55:36:02.65 & 0.0374 & 1133 & Mrk 520 & 22:00:41.382 & +10:33:07.94 & 0.0275 \\
        446 & LEDA 2265450 & 09:11:29.996 & +45:28:06.03 & 0.0267 & 1141 & NGC 7214 & 22:09:07.690 & $-$27:48:34.03 & 0.0227 \\
        449 & Mrk 704 & 09:18:25.979 & +16:18:19.67 & 0.0295 & 1150 & LEDA 68747 & 22:23:45.022 & +11:50:08.63 & 0.0296 \\
        450 & LEDA101470 & 09:19:13.228 & +55:27:55.21 & 0.0490 & 1156 & LEDA 69216 & 22:34:49.827 & $-$25:40:37.07 & 0.0265 \\
        451 & IC 2461 & 09:19:58.030 & +37:11:27.77 & 0.0075 & 1157 & NGC 7314 & 22:35:46.230 & $-$26:03:00.90 & 0.0046 \\
        453 & LEDA 26440 & 09:20:46.242 & $-$08:03:22.41 & 0.0196 & 1161 & Mrk 915 & 22:36:46.487 & $-$12:32:42.63 & 0.0240 \\
        455 & LEDA 26614 & 09:23:43.008 & +22:54:32.44 & 0.0333 & 1162 & LEDA 69449 & 22:40:17.088 & +08:03:13.41 & 0.0250 \\
        458 & Mrk 110 & 09:25:12.871 & +52:17:10.50 & 0.0355 & 1177 & LEDA 70163 & 22:58:55.283 & +40:55:55.97 & 0.0172 \\
        459 & SWIFT J0926.1+6931 & 09:25:47.505 & +69:27:53.27 & 0.0400 & 1178 & LEDA 70195 & 22:59:32.952 & +24:55:05.74 & 0.0333 \\
        460 & Mrk 705 & 09:26:03.245 & +12:44:04.11 & 0.0289 & 1183 & Mrk 926 & 23:04:43.499 & $-$08:41:08.46 & 0.0468 \\
        461 & NGC 2885 & 09:27:18.490 & +23:01:12.16 & 0.0268 & 1185 & LEDA 70504 & 23:07:02.900 & +04:32:56.76 & 0.0406 \\
        463 & LEDA 26940 & 09:29:37.910 & +62:32:38.26 & 0.0256 & 1186 & LEDA 70537 & 23:07:48.875 & +22:42:36.79 & 0.0347 \\
        470 & LEDA 27720 & 09:42:04.768 & +23:41:06.62 & 0.0217 & 1189 & NGC 7603 & 23:18:56.638 & +00:14:37.63 & 0.0293 \\
        475 & NGC 3035 & 09:51:55.027 & $-$06:49:22.50 & 0.0144 & 1192 & LEDA 1029085 & 23:22:24.441 & $-$06:45:37.55 & 0.0330 \\
        481 & NGC 3080 & 09:59:55.880 & +13:02:38.17 & 0.0356 & 1198 & NGC 7682 & 23:29:03.896 & +03:32:00.00 & 0.0167 \\
        486 & LEDA 29323 & 10:05:55.371 & $-$23:03:25.11 & 0.0129 & 1199 & LEDA 2742800 & 23:30:37.712 & +71:22:46.42 & 0.0369 \\
        493 & LEDA 1063109 & 10:17:16.825 & $-$04:04:55.88 & 0.0409 & 1202 & LEDA 72148 & 23:41:55.452 & +30:34:54.23 & 0.0174 \\
        496 & LEDA 30311 & 10:21:40.228 & $-$03:27:14.27 & 0.0410 & & & & & \\
		\hline
		\hline
	\end{tabular}
	}
	\begin{flushleft}
	Note. (1) Entry number listed in the {\it Swift} BAT 70-month hard X-ray survey \citep{Baumgartner2013}; (2) Source name; (3), (4), and (5) Right ascension and declination in J2000 coordinate, and redshift obtained from BASS DR1 \citep{Koss2017}.
	\end{flushleft}
\end{table*}



\bsp	
\label{lastpage}
\end{document}